\documentclass[iop,apjl]{emulateapj}

\usepackage{amsmath}

\shorttitle{Two Satellites of Andromeda}
\shortauthors{Slater et al.}

\slugcomment{}

\begin{document}

\title{A Deep Study of the Dwarf Satellites Andromeda XXVIII \& Andromeda XXIX}

\author{Colin T. Slater and Eric F. Bell}
\affil{Department of Astronomy, University of Michigan,
    1085 S. University Ave., Ann Arbor, MI 48109, USA}

\author{Nicolas F. Martin}
\affil{Observatoire astronomique de Strasbourg, Universit\'e de
Strasbourg, CNRS, UMR 7550, 11 rue de l'Universit\'e, F-67000 Strasbourg,
France}

\author{Erik J. Tollerud and Nhung Ho}
\affil{Astronomy Department, Yale University, P.O. Box 208101, New Haven, CT
06510, USA}

\begin{abstract}
We present the results of a deep study of the isolated dwarf galaxies Andromeda
XXVIII and Andromeda XXIX with Gemini/GMOS and Keck/DEIMOS. Both galaxies are
shown to host old, metal-poor stellar populations with no detectable recent star
formation, conclusively identifying both of them as dwarf spheroidal galaxies
(dSphs). And XXVIII exhibits a complex horizontal branch morphology, which is
suggestive of metallicity enrichment and thus an extended period of star
formation in the past. Decomposing the horizontal branch into blue (metal poor,
assumed to be older) and red (relatively more metal rich, assumed to be younger)
populations shows that the metal rich are also more spatially concentrated in
the center of the galaxy. We use spectroscopic measurements of the Calcium
triplet, combined with the improved precision of the Gemini photometry, to
measure the metallicity of the galaxies, confirming the metallicity spread and
showing that they both lie on the luminosity-metallicity relation for dwarf
satellites. Taken together, the galaxies exhibit largely typical properties for
dSphs despite their significant distances from M31. These dwarfs thus
place particularly significant constraints on models of dSph formation involving
environmental processes such as tidal or ram pressure stripping. Such models
must be able to completely transform the two galaxies into dSphs in no more than
two pericentric passages around M31, while maintaining a significant stellar
populations gradient. Reproducing these features is a prime requirement for
models of dSph formation to demonstrate not just the plausibility of
environmental transformation but the capability of accurately recreating real
dSphs.
\end{abstract}


\section{Introduction}

The unique physical properties and environments of dwarf galaxies make them
excellent test cases for improving our understanding of the processes that
affect the structure, stellar populations, and evolution of galaxies. Because of
their shallow potential wells, dwarf galaxies are particularly sensitive to a
wide range of processes that may only weakly affect larger galaxies. These
processes range from cosmological scales, such as heating by the UV background
radiation \citep{gnedin00}, to interactions at galaxy scales such as tidal
stripping and tidal stirring \citep{mayer01,klimentowski09,kravtsov04}, resonant
stripping \citep{donghia09}, and ram pressure stripping \citep{mayer06}, to the
effects of feedback from from the dwarfs themselves
\citep{dekel86,maclow99,gnedin02,sawala10}. 

Many studies have focused on understanding the differences between the gas-rich,
star forming dwarf irregular galaxies (dIrrs) and the gas-poor, non-star-forming
dwarf spheroidals. While a number of processes could suitably recreate
the broad properties of this differentiation, finding observational evidence in
support of any specific theory has been difficult. One of the main clues in this
effort is the spatial distribution of dwarfs; while dIrrs can be found
throughout the Local Group, dSphs principally are only found within 200-300 kpc
of a larger host galaxy such as the Milky Way or Andromeda
\citep{einasto74,vandenbergh94,grebel03}. This is trend is also reflected in the
gas content of Local Group dwarfs \citep{blitz00,grcevich09}. This spatial
dependence seems to indicate that environmental effects such as tides and ram
pressure stripping are likely to be responsible for creating dSphs. However,
there are outliers from this trend, such as Cetus, Tucana, and Andromeda XV,
which are dSphs that lie more than 700 kpc from either the Milky Way or
Andromeda. The existence of such distant dSphs may suggest that alternative
channels for dSph formation exist \citep{kazantzidis11b}, or it could be an
incidental effect seen in galaxies that have passed through a larger host on
very radial orbits \citep{teyssier12,slater13}.

The set of isolated dwarf galaxies was recently enlarged by the discovery of
Andromeda XXVIII and XXIX, which by their position on the sky were known to be
approximately 360 and 200 kpc from Andromeda, respectively
\citep{slater11,bell11}. While And XXIX was identified as a dSph by the images
confirming it as a galaxy, there was no comparable data on And XXVIII (beyond
the initial SDSS discovery data) with which to identify it as a dSph or dIrr. We
thus sought to obtain deeper imaging of both galaxies down to the horizontal
branch level which would enable a conclusive identification of the galaxies as
dSphs or dIrrs by constraining any possible recent star formation. In addition,
the deep photometry permits more precise determination of the spatial structure
and enables the interpretation of the spectroscopic Calcium triplet
data from \citet{tollerud13} to obtain a metallicity measurement. As we will
discuss, the information derived from these measurements along with dynamical
considerations imposed by their position in the Local Group can together place
significant constraints on plausible mechanisms for the origin of these two
dSphs.

This work is organized as follows: we discuss the imaging data and the reduction process
in Section~\ref{data}, and illustrate the general features of the
color-magnitude diagram in Section~\ref{obsCMD}. Spectroscopic metallicities are
presented in Section~\ref{spectra}, and the structure and stellar populations of
the dwarfs are discussed in Section~\ref{structure}. We discuss the implications
of these results for theories of dSph formation in Section~\ref{discussion}.

\section{Imaging Observations \& Data Reduction}
\label{data}

Between 22 July 2012 and 13 August 2012 we obtained deep images of And XXVIII
and XXIX with the GMOS instrument on Gemini-North (Gemini program
GN-2012B-Q-40). The observations for each dwarf consisted of a total of 3150
seconds in SDSS-i band and 2925 seconds in r, centered on the dwarf. Because the
dwarfs each nearly fill the field of view of the instrument, we also obtained a
pair of flanking exposures for each dwarf to provide an ``off-source'' region for
estimating the contamination from background sources. These exposures
consisted of at least 1350 s in both r and i, though some fields received a
small number of extra exposures. The images were all taken in 70th percentile
image quality conditions or better, which yielded excellent results with the
point source full width at half maximum ranging between 0.47\arcsec and
0.8\arcsec. 

\begin{figure*}
\epsscale{1.0}
\plotone{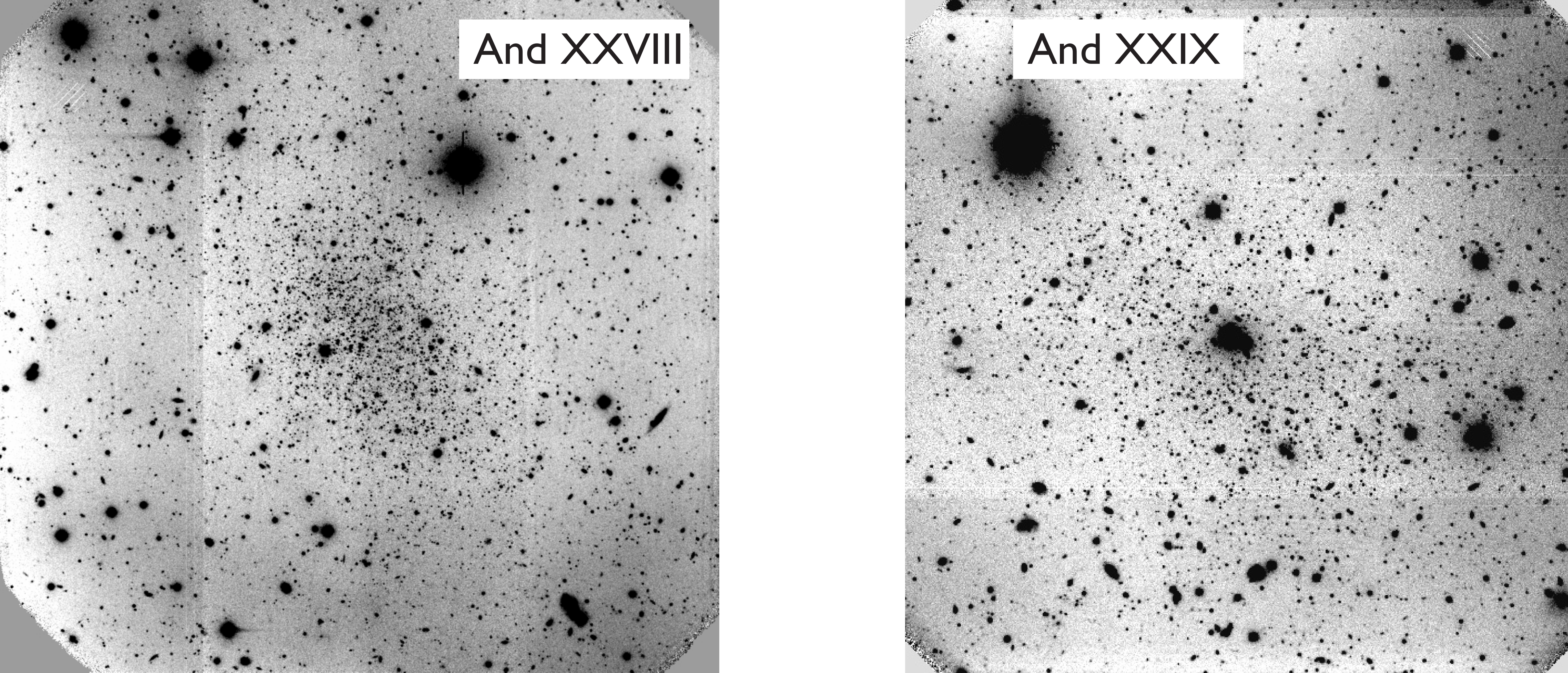}
\caption{Stacked i-band image of And XXVIII on the left, and of And XXIX on the
right. North is up, and East is to the left. Both images are approximately
5.6\arcmin on a side. The saturated feature near the center of And XXIX is a
combination of a foreground star and two background galaxies. \label{images}}
\end{figure*}

All of the images were bias subtracted, flat fielded, and coadded using
the standard bias frames and twilight flats provided by Gemini. The
reduced images can be seen in Figure~\ref{images}. Residual flat fielding
and/or background subtraction uncertainty exists at the 1\% level (0.01
magnitudes, roughly peak to valley). PSF photometry
was performed using DAOPHOT \citep{stetson87}, which enabled accurate
measurements even in the somewhat crowded centers of the dwarfs. In many cases
the seeing in one filter was much better than the other, such as for the core of
And XXVIII where the seeing was $0.47\arcsec$ in i and $0.68\arcsec$ in r. In these
cases we chose to first detect and measure the position of stars in the image
with the best seeing, and then require the photometry of the other band to reuse
the positions of stars detected in the better band. This significantly extends
our detection limit, which would otherwise be set by the shallower band, but
with limited color information at these faint magnitudes.

The images were calibrated to measurements from the Sloan Digital Sky Survey
(SDSS), Data Release 9 \citep{DR9}. For each stacked image we cross-matched all
objects from the SDSS catalog that overlapped our fields, with colors between
$-0.2 < (r-i)_0 < 0.6$, and classified as stars both by SDSS and DAOPHOT.
Star-galaxy separation was performed using the ``sharp'' parameter from DAOPHOT.
From this we measured the weighted mean offset between the SDSS magnitudes and
the instrumental magnitudes to determine the zeropoint for each field. 
Between the saturation limit of the Gemini data, mitigated by taking several
exposures, and faint limits of the SDSS data (corresponding to approximately $19
< i < 22.5$ and $19.5 < r < 22.5$) there were of order 100 stars used for the
calibration of each frame. Based on the calculated stellar measurement
uncertainties the formal uncertainty on the calibration is at the millimagnitude
level, but unaccounted systematic effects likely dominate the statistical
uncertainty (e.g., precision reddening measurements). All magnitudes were
dereddened with the extinction values from \citet{schlafly11}.

The photometric completeness of each stacked image was estimated by artificial
star tests. For each field we took the PSF used by DAOPHOT for that field and
inserted a large grid of artificial stars, with all of the stars at the same
magnitude but with Poisson noise on the actual pixel values added to each
image. This was performed for both r and i band images simultaneously, and the
resulting pair of images was then run through the same automated DAOPHOT
pipeline that was used on the original image. Artificial stars were inserted
over a a grid of i band magnitudes and r-i colors, producing measurements of the
recovery rate that cover the entire CMD. The 50\% completeness limit for both
dwarfs is at least $r_0 = 25.5$, with slightly deeper data in the i-band for And
XXVIII.

The observed CMDs suffer from both foreground and background contamination.
Foreground dwarf stars in the Milky Way tend to contribute at the bright end of
the CMD. At the faint end, distant galaxies that are too small to be resolved
become the dominant source of contamination. This effect can quickly become
significant at fainter magnitudes due to the rapid rise in the observed galaxy
luminosity function. This effect was minimized by the superb seeing at the
Gemini observatory, which allowed smaller galaxies to be resolved and excluded
from our sample.

\section{Observed CMDs}
\label{obsCMD}

\begin{figure*}
\epsscale{0.8}
\plotone{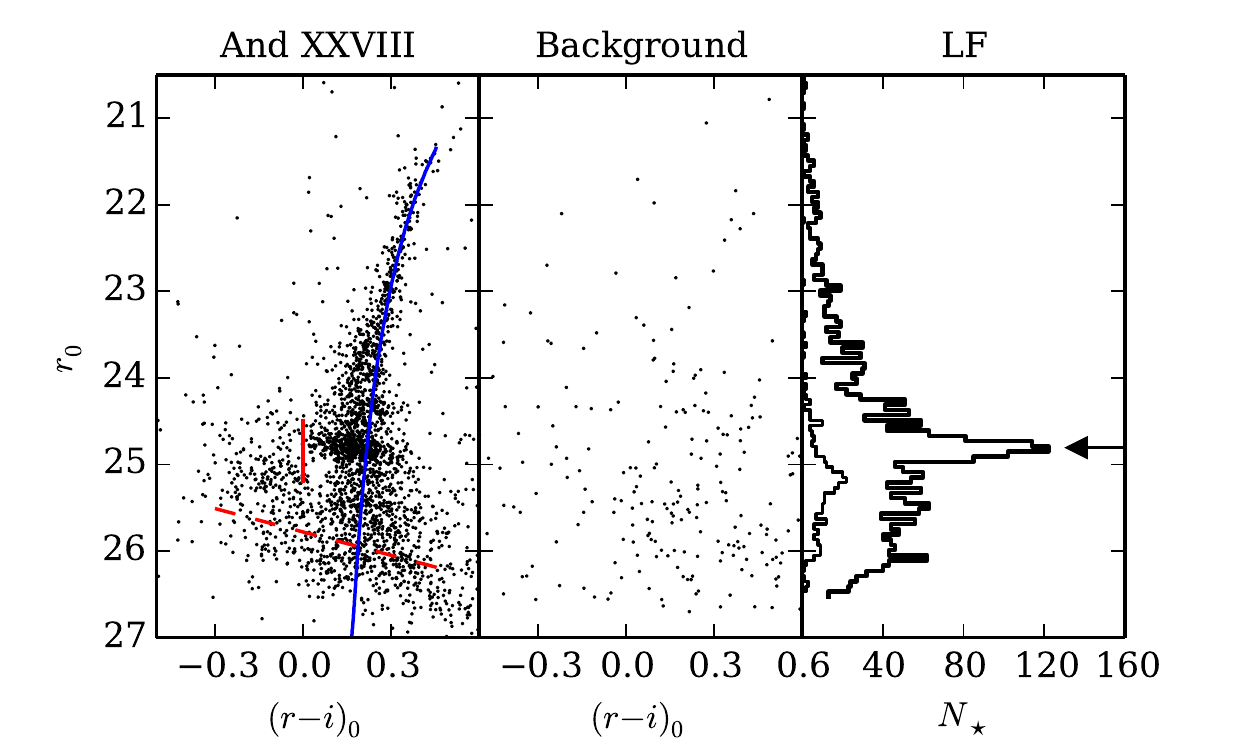}
\caption{CMD of And XXVIII on the left (inside $2r_h$), with the CMD of an
equal-sized background region in the center. The red dashed line
indicates the 50\% completeness limit, while the vertical red line indicates the
approximate division between red and blue horizontal branches. The luminosity
function of the dwarf is shown on the right, separated into a thick line showing
stars with $(r-i)_0 > 0$ and a thin line showing stars with $(r-i)_0 < 0$. A 12
Gyr old, [Fe/H] = -1.84 isochrone is overplotted, and the measured apparent
magnitude of the HB is indicated with an arrow. \label{cmd_28}}
\end{figure*}

\begin{figure*}
\epsscale{0.8}
\plotone{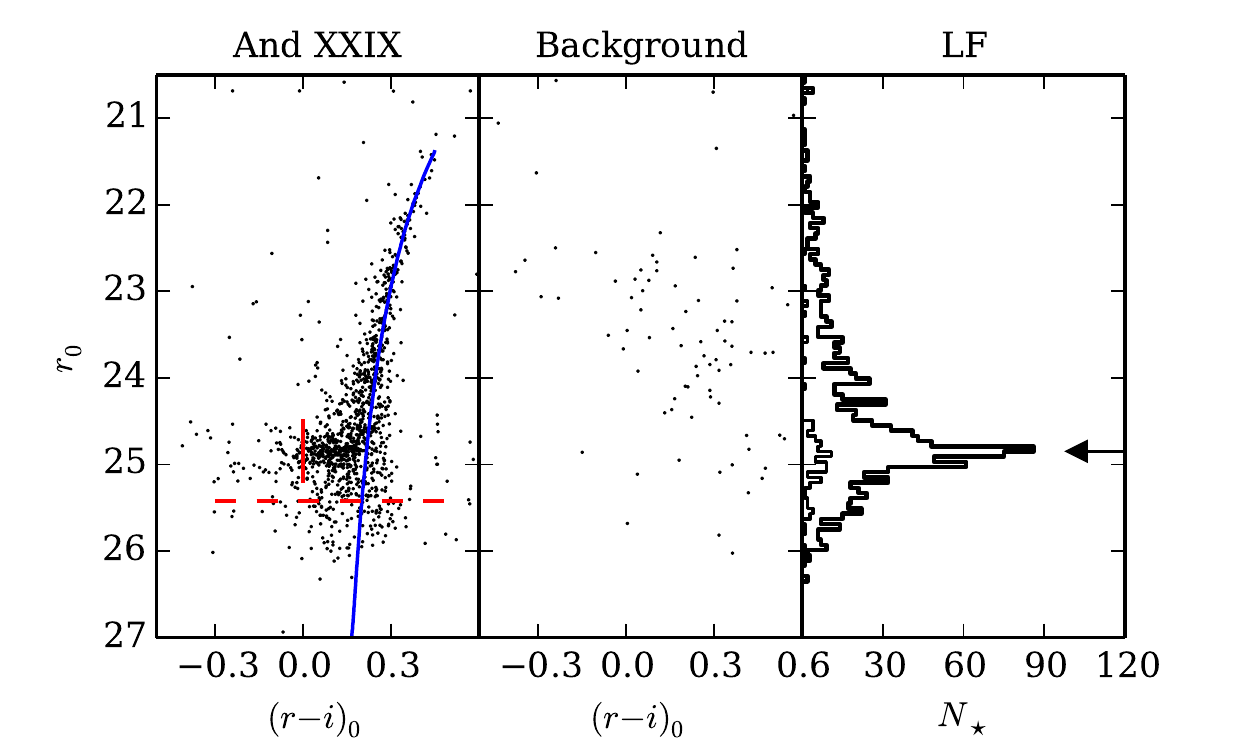}
\caption{The same panels as Figure~\ref{cmd_28}, but for And XXIX. A 12 Gyr old,
[Fe/H] = -1.92 isochrone is overplotted. As with And XXVIII there are no
indications of recent star formation. Though there may be some hints of a
BHB, if it does exist it is substantially less prominent than in And XXVIII.
\label{cmd_29}}
\end{figure*}

The CMDs of And XXVIII and XXIX are shown in the left panels of
Figures~\ref{cmd_28} and \ref{cmd_29}, respectively. A 12 Gyr old isochrone from
\citet{dotter08} is overlaid at the distances and spectroscopic metallicities
determined later in this work. Both dwarfs show a
well-populated giant branch with a very prominent red clump/red horizontal
branch (RC/RHB) near $r_0 \sim 24.5$ - $25.0$. This feature is particularly
clear as a large bump in the luminosity functions of each dwarf, shown by the
thick black line in the right panels of Figure~\ref{cmd_28} and \ref{cmd_29}. In
addition to the RC/RHB, And XXVIII also shows a blue horizontal branch (BHB)
slightly fainter than $r_0 \sim 25.0$ and spanning $-0.3 < (r-i)_0 < 0.0$ in
color. The luminosity function for stars with $(r-i)_0 < 0.0$ is shown by the
thin line on the right panel of Figure~\ref{cmd_28}. The presence of a complex
horizontal branch suggests that And XXVIII has had an extended star formation
history (SFH), since the BHB is typically seen in the oldest globular clusters,
while the RHB tends to appear in globular clusters roughly 2-4 Gyr younger than
the oldest populations \citep{stetson89,sarajedini95}, although a few
globular clusters do show both BHB and RHB \citep{an08}. The additional
information from the spectroscopic metallicity spread, as will be discussed
below, also confirms the extended star formation in both dwarfs.
And XXIX does not show the same prominent BHB. There are 5-10 stars in a similar
position as the BHB in And XXVIII, but this is almost negligible compared to the
100 or more stars in the BHB of And XXVIII and could be background
contamination. This does not indicate that there is no ancient population in And
XXIX, as, for example, the Draco dSph also contains very few BHB stars
\citep{segall07}.

There is a notable absence of any young main-sequence stars in the observed CMDs
of both XXVIII and XXIX, which suggests that there has not been any recent star
formation at appreciable rates in either galaxy. The handful of stars brighter
than the HB and on the blue side of the RGB are consistent with foreground (or
background) contamination. The CMD of And XXIX has an almost negligible number
of stars bluewards of the RGB at any magnitude. The CMD of And XXVIII does show
some blue detections below the BHB, but it is difficult to conclusively identify
their origin. Since the precision of the colors degrades at faint magnitudes,
these detections could be an (artificial) broadening of the RGB, possibly
scattering more stars towards the blue due to the somewhat shallower depth of
the r-band exposures. It is also possible that they are background sources or
false detections from noise, both of which could be strongly weighted towards
the faintest magnitudes. None of these origins are clearly favored and some
combination could be at work, but there is not sufficient evidence to believe
that these sources are main sequence stars.

The absence of observed young main sequence stars in And XXVIII is complimented by
recent work that shows little to no cold gas in the galaxy. Observations with
the Westerbork Synthesis Radio Telescope place a $5-\sigma$ upper limit on the
total HI mass of $2.8 \times 10^3$ $M_\odot$ (T. Oosterloo, private
communication). For comparison, the similarly low-mass dwarf Leo T has had
recent star formation and contains $\sim 2.8 \times 10^5$ $M_\odot$ of
H\textsc{I} \citep{ryan-weber08}, while most dSphs have upper limits at this
level or less \citep{grcevich09}. This stringent limit on the gas in And XXVIII
adds further evidence that it is a dSph. 

\begin{table}
\caption{Properties of And XXVIII \& XXIX\label{properties_table}}
\begin{center}
\begin{tabular}{l c c}
\tableline
Parameter & And XXVIII & And XXIX\\
\tableline
\tableline
$\alpha$ (J2000) &  $22^{\rm h}$ $32^{\rm m}$ $41\fs 5$ &  $23^{\rm h}$ $58^{\rm m}$ $55\fs 6$\\
$\delta$ (J2000) &  $31\arcdeg$ $13\arcmin$ $3.7\arcsec$ & $30\arcdeg$ $45\arcmin$ $20.2\arcsec$\\
E(B-V) & 0.080\tablenotemark{a} & 0.040\tablenotemark{a}  \\
Ellipticity & $0.43$ $\pm$ $0.02$ & $0.29$ $\pm$ $0.04$\\
Position Angle (N to E) & $34^\circ$ $\pm$ $1^\circ$ & $55^\circ$ $\pm$ $4^\circ$\\
$r_h$ & $1\farcm 20$ $\pm$ $0\farcm 03$  & $1\farcm 39$ $\pm$ $0\farcm 08$ \\
$r_h$ & $280 \pm 20$ pc & $315 \pm 15$ pc\\
$D$ & $811$ $\pm$ $48$ kpc & $829$ $\pm$ $42$ kpc \\
$(m - M)_0$ & $24.55 \pm 0.13$ & $24.59 \pm 0.11$\\
$r_{\rm M31}$\tablenotemark{b} & $385^{+18}_{-13}$ kpc & $198^{+18}_{-10}$ kpc\\
$M_V$ & $-8.7\pm0.4$ & $-8.5\pm0.3$\\
$\langle[{\rm Fe/H}]\rangle$ & $-1.84 \pm 0.15$ & $-1.90 \pm 0.12$\\
$\sigma([{\rm Fe/H}])$ & $0.65 \pm 0.15$ & $0.57 \pm 0.11$\\
HI & $< 2.8 \times 10^3 M_\odot$ & \\

\tableline
\tablenotetext{1}{\citet{schlafly11}}
\tablenotetext{2}{3D distance, rather than projected.}
\end{tabular}
\end{center}
\end{table}

\subsection{Distance and Luminosity}

The clear HB in both dwarfs enables an accurate measurement of the distance to
the dwarfs, and hence their distance to M31. We fit a Gaussian plus a linear
background model to the r-band luminosity function of each dwarf in the region
of the HB, using only stars redder than $(r-i)_0 = 0$. The measured HB position
is indicated in the right panels of Figure~\ref{cmd_28} and \ref{cmd_29} by the
horizontal arrow, and is $m_{g,0} = 24.81$ for And XXVIII and $m_{g,0} = 24.84$
for And XXIX. We use the RHB absolute magnitude calibration of \citet{chen09},
which is based on globular clusters RHBs measured directly in the SDSS filter
set. In the r-band this calibration, using a linear metallicity dependence and
without the age term, is
\begin{equation}
M_r = 0.165 [\textrm{Fe/H}] + 0.569.
\end{equation}

The resulting distances are $811 \pm 48$ kpc for And XXVIII
and $829 \pm 42 $kpc for And XXIX, using the spectroscopic metallicities as
determined in Section~\ref{spectra}. Both of these are slightly further than the
measured distances from \citet{slater11} and \citet{bell11}, but just within
(And XXVIII) or just outside (And XXIX) the formal one-sigma uncertainties.  The
updated heliocentric distances does not substantially change the measured
distances between the dwarfs and M31, since both are near the tangent point
relative to M31\footnote{The distance between And XXIX and M31 reported in
\citet{bell11} was incorrect due to a geometry error; it is fixed in this
work.}. Based on these distances, both dwarfs lie well away from the
plane of satellites from \citet{conn13} and \citet{ibata13}. As seen from M31
the satellites are $80^\circ$ (And XXVIII) and $60^\circ$ (And XXIX) from the
plane. The closest galaxy to And XXVIII is And XXXI at 164 kpc, while And XXIX's
closest neighbor is And XIX at 88 kpc, making both relatively isolated from
other dwarfs.

We measured the total luminosity of both dwarfs by comparing the portion of the
LF brighter than the HB to the LF of the Draco dwarf. Using data from
\citet{segall07} we constructed a background-subtracted LF for Draco inside
$r_h$, then scaled the LF of the dwarfs such that they best matched the Draco
LF. The resulting luminosities are $M_V = -8.7 \pm 0.4$ for And XXVIII and $M_V
= -8.5 \pm 0.3$ for And XXIX, both of which are again in good agreement with
values measured by previous works.

\section{Spectroscopic Metallicity}
\label{spectra}

\begin{figure}
\epsscale{1.2}
\plotone{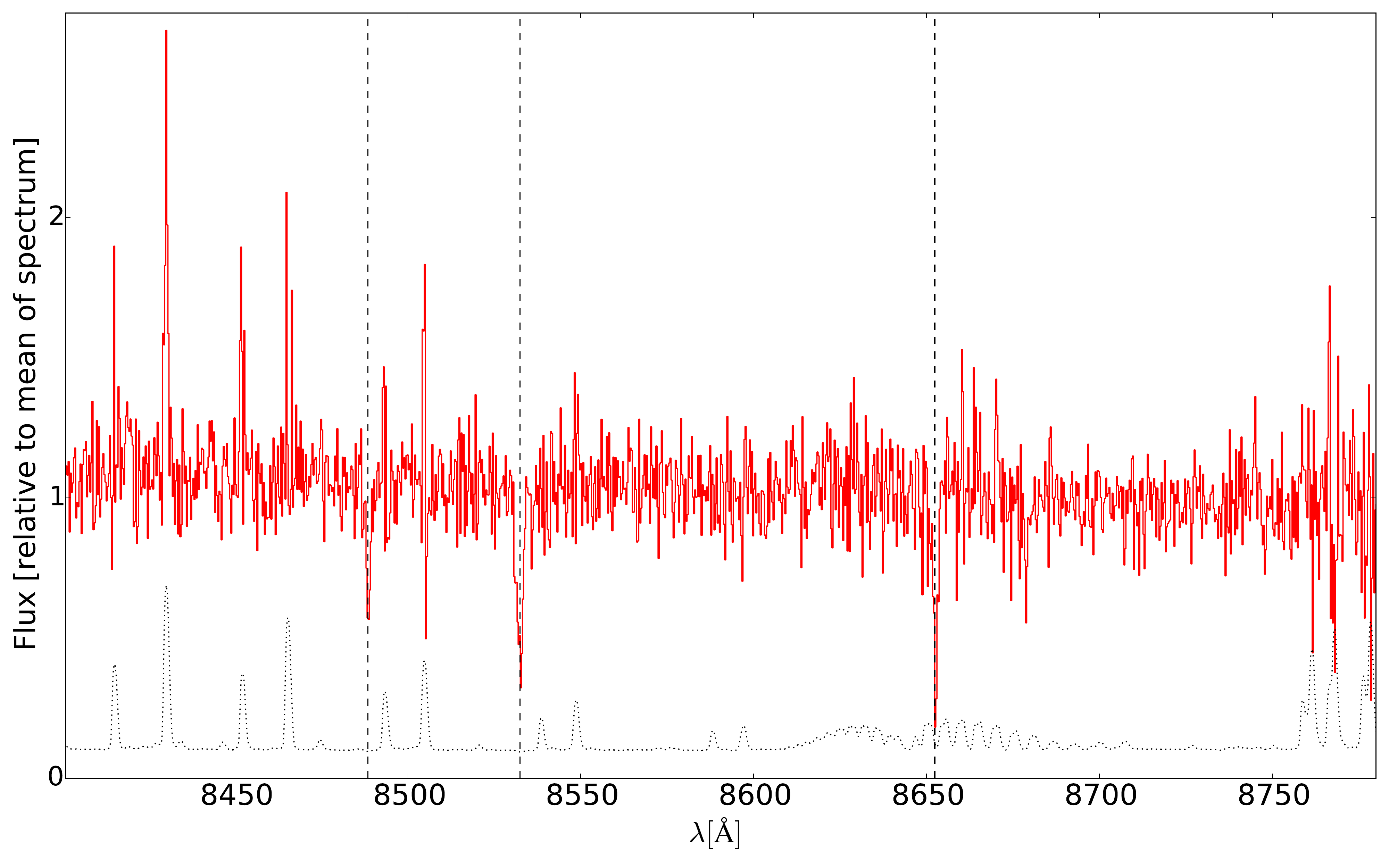}
\caption{An example spectrum of an individual star in And XXVIII, focusing on the
triplet of Calcium lines (marked with vertical dashed lines, and shifted to the
velocity of And XXVIII). The dotted line indicates the RMS uncertainty at each
point in the spectrum. \label{sample_spectrum}}
\end{figure}

To complement the imaging data, we also make use of metallicities derived from
Keck/DEIMOS spectroscopy of the brightest RGB stars. The source data and
spectroscopic reductions are described in \citet{tollerud13}, and a
sample spectrum is shown in Figure~\ref{sample_spectrum}. We derive
metallicities from the $\lambda \sim 8550$ \AA{} Calcium triplet features,
following the methodology described in \citet{ho13}. Briefly, this procedure
fits a Gaussian profiles to the strongest two CaT lines, and uses these fits to
derive CaT equivalent widths. In combination with absolute magnitudes from the
aforementioned photometric data (Section~\ref{data}), these data can be
calibrated to act as effective proxies for $[{\rm Fe/H}]$ of these stars. For
this purpose, we adopt the \citet{carrera13} metallicity calibration to convert
our photometry and equivalent widths to $[{\rm Fe/H}]$.

A table of the spectroscopic metallicity measurements of individual stars in
each dwarf is presented in Table~\ref{stars_table}. We determine the uncertainty
in the galaxy mean [Fe/H] by performing 1000 Monte Carlo resamplings of the
distribution. For each resampling, we add a random offset to the metallicity of
each star drawn from a Gaussian with width of the per-star [Fe/H] uncertainty,
and compute the mean of the resulting distribution. For measuring each galaxy's
metallicity spread $\sigma({\rm[Fe/H]})$, we report the second moment of the
individual measurement distribution and derive uncertainties from a resampling
procedure like that for the galaxy mean [Fe/H]. 

\begin{figure}
\epsscale{1.2}
\plotone{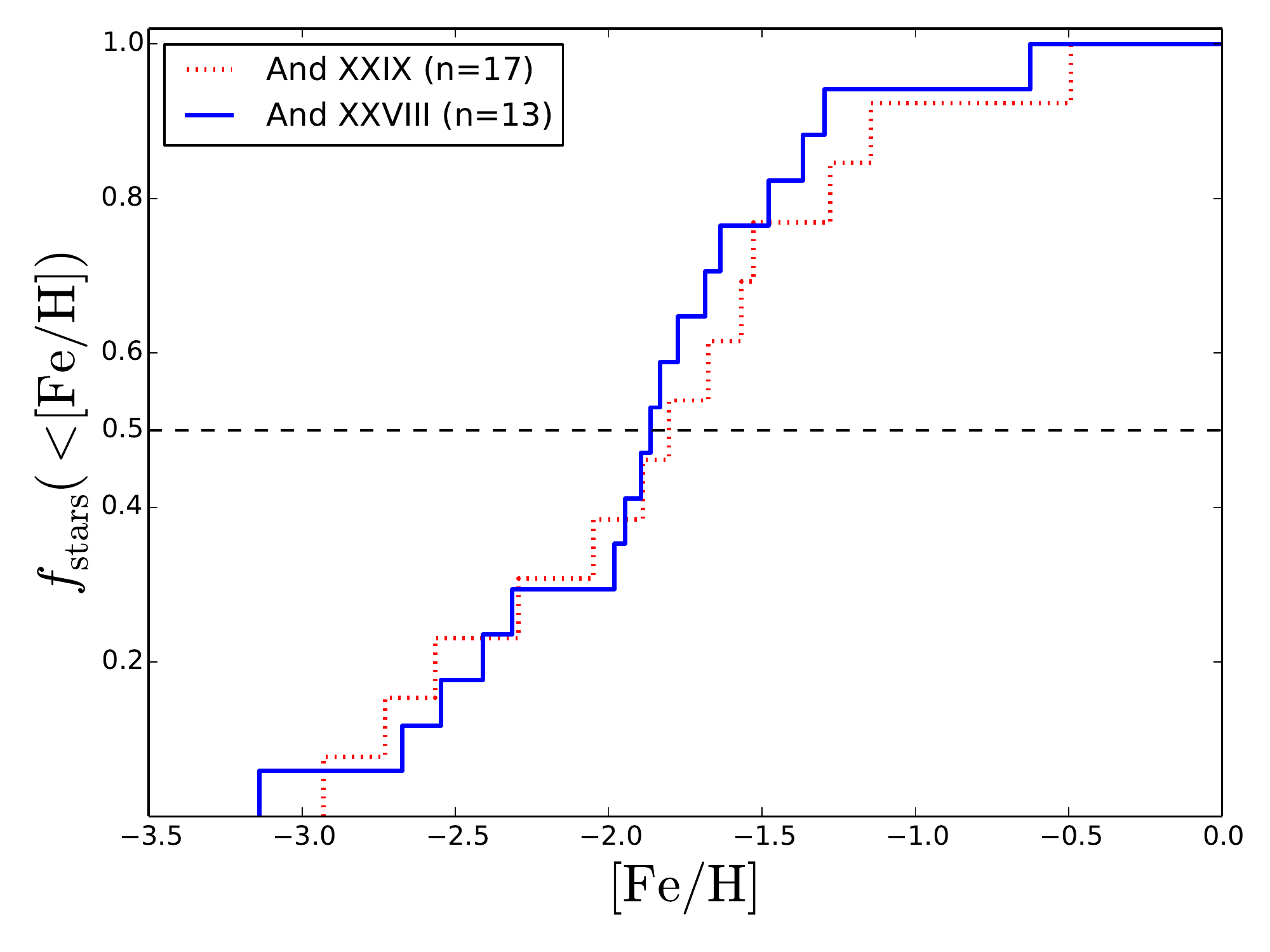}
\caption{Cumulative distribution of $[{\rm Fe/H}]$ for And XXVIII (blue solid
line) and XXIX (red dotted line). \label{fehcdfs}}
\end{figure}

\begin{figure}
\epsscale{1.2}
\plotone{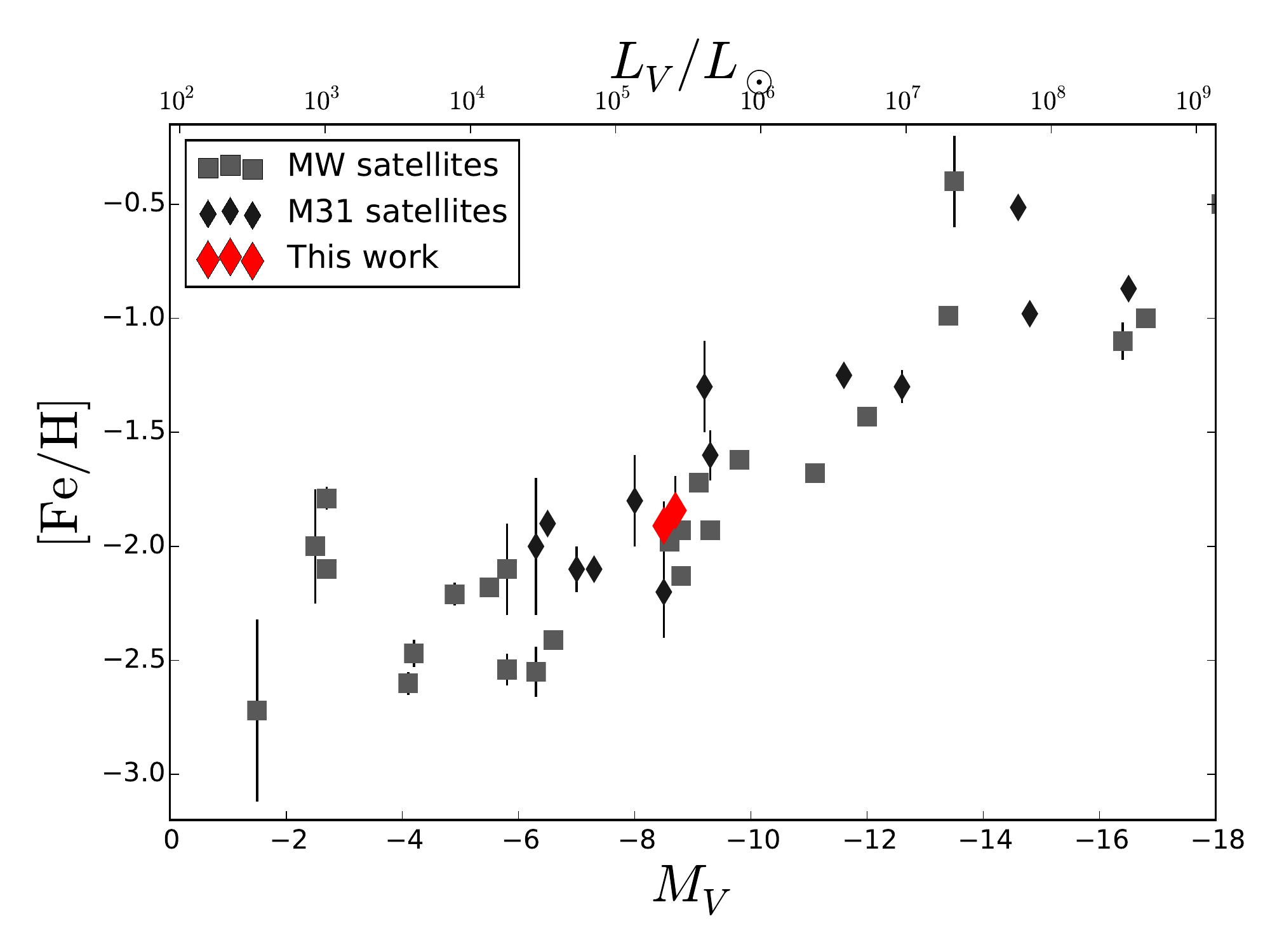}  
\caption{Luminosity-metallicity relation for Local Group satellites, adapted
from the compilation presented in \citet{ho13}, which includes data from
\citet{kirby11} and \citet{collins13}. Squares are MW satellites, diamonds are
M31 satellites, and the error bars are from the Monte Carlo resampling of the
[Fe/H] distribution for each galaxy. And XXVIII and XXIX are shown as the larger
red diamonds. This demonstrates that And XXVIII and XXIX lie on the same
metallicity-luminosity relation as other Local Group satellites.
\label{lfeh}}
\end{figure}

The resulting metallicity distributions for And XXVIII and XXIX are shown as
cumulative distribution functions in Figure \ref{fehcdfs}. From this it is
immediately clear that, while the number of stars are relatively small, the
median of the distribution peaks at $[{\rm Fe/H}] \sim -2$ (see
Table~\ref{properties_table}). Motivated by this, in Figure~\ref{lfeh}, we show
the luminosity-metallicity relation for the brighter M31 satellites \citep{ho13}
and the MW satellites \citep{kirby11,kirby13}, using luminosities from Martin et
al. (2015, submitted). It is immediately clear from this figure
that And XXVIII and XXIX are fully consistent with the metallicity-luminosity
relation that holds for other Local Group satellites. Our measurement for And~
XXVIII is also consistent with the prior measurement by \citet{collins13} of
[Fe/H] $= -2.1 \pm 0.3$, but at higher precision.

\begin{table*}
\begin{center}
\caption{Metallicities of And XXVIII \& XXIX Stars\label{stars_table}}
\begin{tabular}{ccccccc}
\hline
Galaxy & RA (deg) & Dec (deg) & $r_0$ & $(r-i)_0$ & [Fe/H] &
$\sigma_{\rm[Fe/H]}$\tablenotemark{a} \\
\hline
And XXVIII & 338.16549 & 31.20840 & 21.397 & 0.53 & -2.29 & 0.5 \\
And XXVIII & 338.18538 & 31.22404 & 21.404 & 0.44 & -1.58 & 0.5 \\
And XXVIII & 338.15561 & 31.18421 & 21.381 & 0.45 & -1.54 & 0.8 \\
And XXVIII & 338.14847 & 31.15615 & 21.001 & 0.24 & -0.50 & 0.8 \\
And XXVIII & 338.17702 & 31.21802 & 21.548 & 0.42 & -2.06 & 0.5 \\
And XXVIII & 338.17499 & 31.22058 & 21.969 & 0.35 & -2.91 & 0.7 \\
And XXVIII & 338.18206 & 31.21668 & 21.509 & 0.41 & -2.74 & 0.4 \\
And XXVIII & 338.16849 & 31.22444 & 22.344 & 0.42 & -1.90 & 0.6 \\
And XXVIII & 338.18357 & 31.21526 & 21.332 & 0.38 & -1.81 & 0.4 \\
And XXVIII & 338.17542 & 31.23720 & 21.57 & 0.70 & -1.28 & 0.3 \\
And XXVIII & 338.15091 & 31.20916 & 21.578 & 0.46 & -1.68 & 0.2 \\
And XXVIII & 338.18428 & 31.23235 & 21.861 & 0.37 & -1.15 & 0.3 \\
And XXVIII & 338.22622 & 31.21862 & 21.78 & 0.37 & -2.57 & 0.2 \\
And XXIX & 359.73912 & 30.74974 & 22.113 & 0.36 & -1.83 & 0.5 \\
And XXIX & 359.72546 & 30.74484 & 21.467 & 0.45 & -1.94 & 0.3 \\
And XXIX & 359.72690 & 30.76834 & 21.592 & 0.44 & -1.29 & 0.4 \\
And XXIX & 359.74259 & 30.75986 & 22.084 & 0.42 & -0.62 & 0.5 \\
And XXIX & 359.74561 & 30.75100 & 21.854 & 0.39 & -2.40 & 0.4 \\
And XXIX & 359.71503 & 30.74976 & 21.369 & 0.40 & -2.54 & 0.3 \\
And XXIX & 359.71755 & 30.74150 & 21.968 & 0.37 & -3.14 & 0.5 \\
And XXIX & 359.71880 & 30.73644 & 22.003 & 0.40 & -1.36 & 0.4 \\
And XXIX & 359.71957 & 30.76735 & 22.211 & 0.35 & -1.86 & 0.6 \\
And XXIX & 359.75409 & 30.76225 & 21.172 & 0.45 & -1.97 & 0.3 \\
And XXIX & 359.75959 & 30.76464 & 22.111 & 0.36 & -1.77 & 0.6 \\
And XXIX & 359.73776 & 30.80015 & 21.266 & 0.20 & -2.31 & 0.3 \\
And XXIX & 359.73609 & 30.79734 & 22.137 & 0.33 & -1.68 & 0.5 \\
And XXIX & 359.68681 & 30.72895 & 21.959 & 0.36 & -2.68 & 0.6 \\
And XXIX & 359.74074 & 30.76867 & 21.407 & 0.44 & -1.89 & 0.4 \\
And XXIX & 359.74687 & 30.76948 & 21.751 & 0.29 & -1.47 & 0.5 \\
And XXIX & 359.75467 & 30.75391 & 21.752 & 0.37 & -1.63 & 0.5 \\
\hline
\tablenotetext{1}{Individual star [Fe/H] uncertainity; not to be confused with
the overall metallicity spread in Table 1.}
\end{tabular}
\end{center}
\end{table*}

\section{Structure \& Stellar Populations}
\label{structure}

We determined the structural properties of the dwarfs using an updated version
of the maximum likelihood method presented in \citet{martin08}. This method fits
an exponential radial density profile to the galaxies without requiring the data
to be binned, which enables more precise measurements of the structure in
galaxies with only a small number of observed stars. The updated version samples
the parameter space with a Markov Chain Monte Carlo process, and can more easily
account for missing data (Martin et al. 2015, submitted.) This is necessary to
account for the limited field of view of GMOS, which could cause a systematic size
error \citep{munoz12}, as well as the very center of And
XXIX where an inconveniently-located bright foreground star contaminates the
very center of the image and prevents reliable photometry in the surrounding
region.

\begin{figure}
\epsscale{1.2}
\plotone{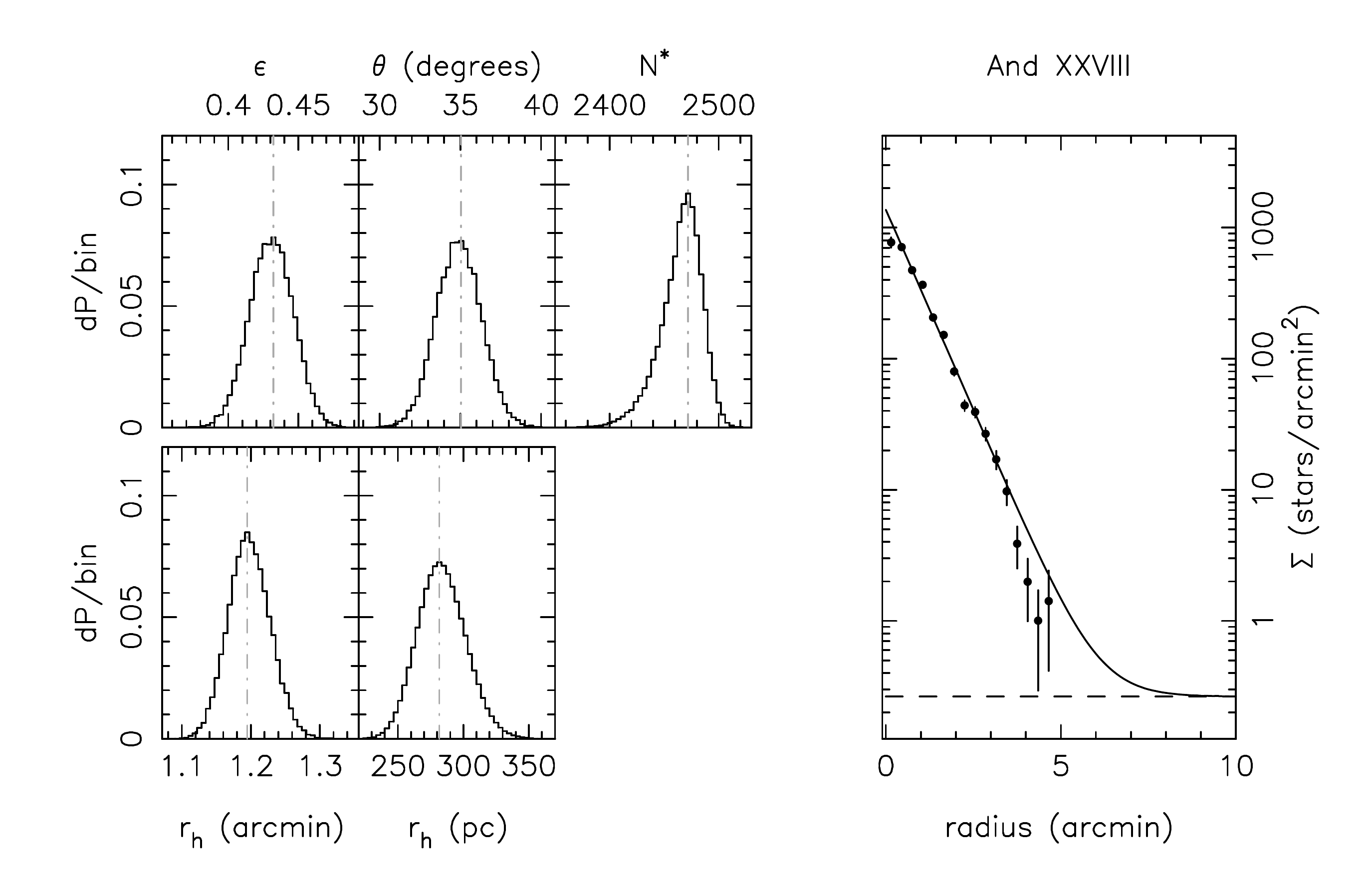}  
\caption{Posterior probability distributions for the structural parameters fit
for And XXVIII are shown on the left. From top-left to bottom-right, these show
these correspond to the ellipticity ($\epsilon$), the position angle from north
to east ($\theta$), the number of stars under the profile for the assumed depth
limit ($N^*$), the angular major-axis half-light radius ($r_h$), and its
corresponding physical length assuming the distance modulus measured above. The
radial profile is shown on the right, with the best fit exponential profile
shown by the solid line and the dashed line showing the background level.
\label{profile_28}}
\end{figure}

\begin{figure}
\epsscale{1.2}
\plotone{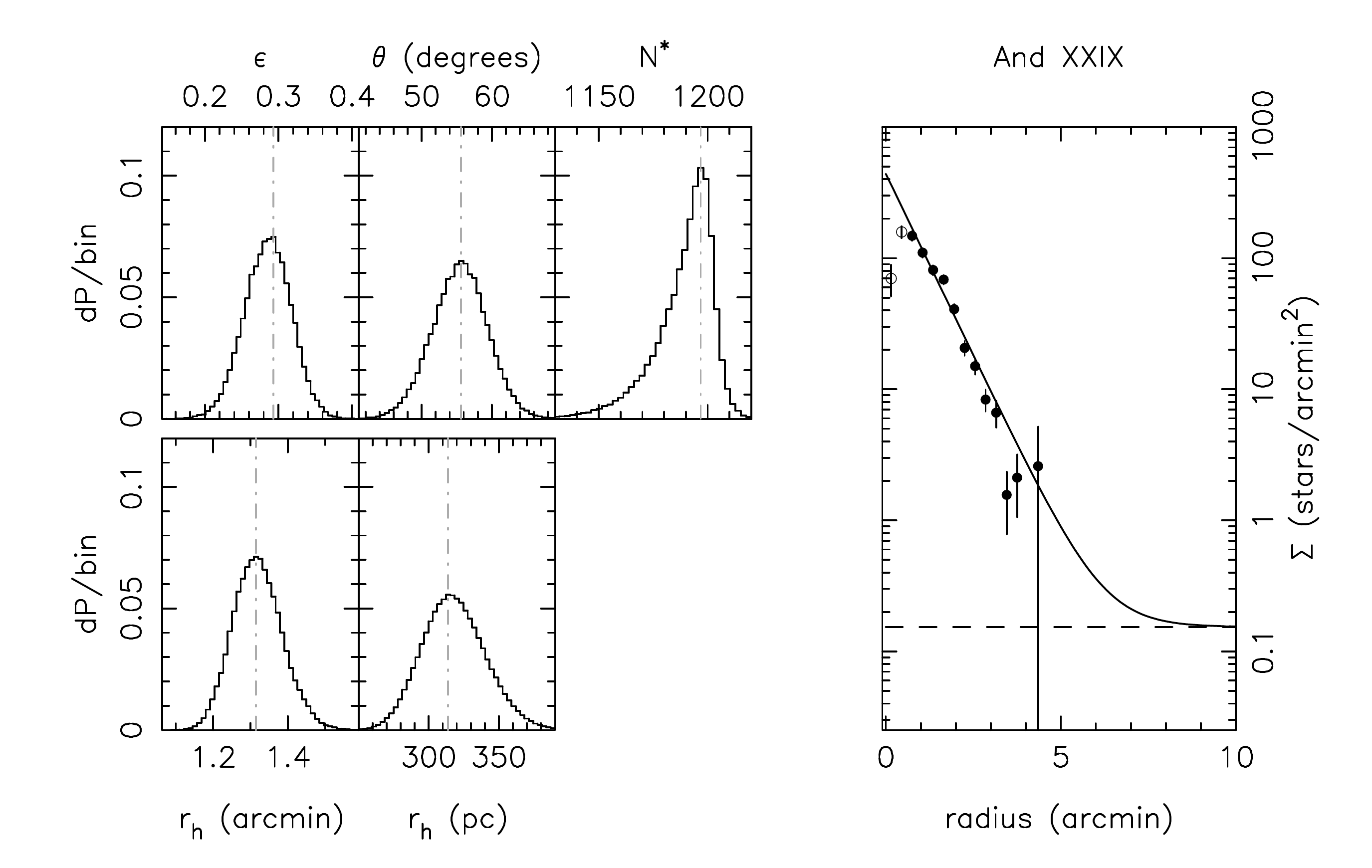}  
\caption{The same posterior probability distributions and radial profile as in
Figure~\ref{profile_28}, but for And XXIX. The two innermost radial profile
points (open circles) were not used in the fit due to the bright contamination
in the center of the galaxy.
\label{profile_29}}
\end{figure}

The resulting radial profiles and posterior probability distributions are shown
in Figures~\ref{profile_28} and \ref{profile_29}. The half-light radii and
ellipticities all have fairly typical values for other dwarfs of similar
luminosities \citep{brasseur11}. The results are also consistent with the
parameters estimated from the much shallower SDSS data \citep{slater11,bell11}.

\begin{figure*}
\epsscale{0.9}
\plotone{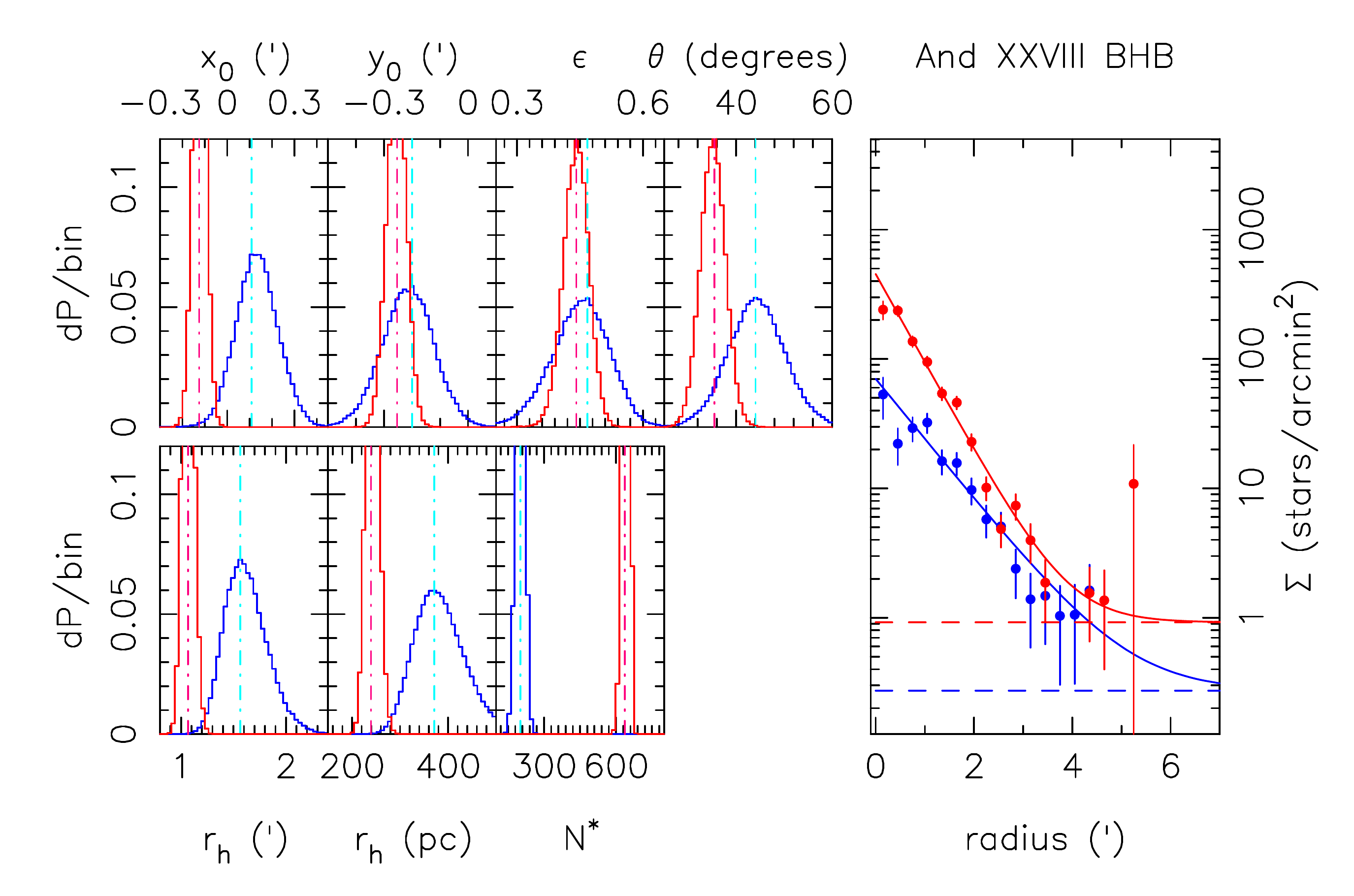}  
\caption{Posterior probability distributions for the structural fit of And
XXVIII, performed separately for stars in the RHB (red lines) and the BHB (blue
lines). The difference in the radial profile clearly visible in the panel on
the right, and the significance is confirmed by the difference in half light
radius ($r_h$). The ellipticities and position angles are similar in the two
populations.
\label{bhb_rhb}}
\end{figure*}


The separation between the red and blue horizontal branches in And XXVIII
enables us to examine the spatial distribution of the metal-poor, older,
and the more metal-rich, younger, stellar populations. Radial profiles of the two
horizontal branches (separated at $(r-i)_0 = 0.0$) are shown in
Figure~\ref{bhb_rhb}. The difference in the radial profiles is easily seen in
the right panel, and the posterior probability distributions for the half-light
radius confirm the statistical significance of the difference. This behavior has
been seen in other dwarf galaxies, such as Sculptor \citep{tolstoy04}, Fornax
\citep{battaglia06}, Canes Venatici I \citep{ibata06}, And II
\citep{mcconnachie07}, and Leo T \citep{dejong08}. In all of these cases the
more metal-rich population is the more centrally concentrated one, consistent
with And XXVIII. Measuring the spatial structure of the two components
independently shows that they appear to be simply scaled versions of each other;
the half-light radii are $370 \pm 60$ pc and $240 \pm 15$ pc (blue and red,
respectively), while the ellipticities of $0.48 \pm 0.06$ and $0.43 \pm 0.03$,
along with position angles of $45^\circ \pm 5^\circ$ and $34^\circ \pm 3^\circ$,
agree well with each other. Taken together this implies that the process that
transformed the dwarf into a pressure-supported system did so without
randomizing the orbital energies of individual stars enough to completely
redistribute the older and younger populations, but both populations did end up
with the same general morphology.

Simulations of isolated dwarfs by \citet{kawata06} are able to reproduce a
radial metallicity gradient, but with some uncertainty over the number of stars
at the lowest metallicity values and the total luminosity of the simulated
dwarfs (and also see \citet{revaz12} for simulated dwarfs without
gradients). In these simulations the metallicity gradient is produced by the
continuous accretion of gas to the center of the galaxy, which tends to cause
more metal enrichment and a younger population (weighted by mass) at small radii
when compared to the outer regions of the galaxy. This explanation suggests that
the ``two populations'' we infer from the RHB and BHB of And XXVIII are perhaps
more properly interpreted as two distinct tracers of what is really a continuous
range of ages and metallicities present in the dwarf. In this scenario, the lack
of observed multiple populations in And XXIX could be the result of the dwarf
lacking sufficient gas accretion and star formation activity to generate a
strong metallicity gradient. If this is the case, then there may be a mass
dependence to the presence of such gradients, which makes it particularly
significant that And XXVIII is a relatively low-mass galaxy to host such a
behavior. Whether this is merely stochasticity, or the influence of
external forces, or if it requires a more complex model of the enrichment
process is an open question.

\section{Discussion and Conclusions}
\label{discussion}

The analysis of And XXVIII and XXIX shows that both galaxies are relatively
typical dwarf spheroidals, with old, metal-poor stellar populations and no
measurable ongoing or recent star formation. The significance of these galaxies
in distinguishing models of dSph formation comes from their considerable
distances from M31. If environment-independent processes such as supernova
feedback or reionization are responsible for transforming dIrrs into dSphs, then
finding dSphs at these distances is quite natural. However, such models are
by themselves largely unable to reproduce the radial dependence of the dSph
distribution around the Milky Way and M31. An environment-based transformation
process, based on some combination of tidal or ram pressure forces, can
potentially account for the radial distribution, but correctly reproducing the
properties of dSphs large radii is the critical test of such models. It is in
this light that Andromeda XXVIII and Andromeda XXIX have the most power to
discriminate between models. 

Models of tidal transformation have been studied extensively and can account for
many of the observed structural properties of dSphs
\citep{mayer01,lokas10,lokas12}. However, a critical component of understanding
whether these models can reproduce the entire population of Local Group dSphs is
the dependence of the transformation process on orbital pericenter distances and
the number of pericentric passages. At large radii the weaker tidal force may
lose its ability completely transform satellites into dSphs, potentially leaving
observable signatures in satellites on the outskirts of host galaxies.

Observationally we cannot directly know the orbital history of individual
satellites without proper motions (of which there are very few), and must test
the radial distribution of dSphs in a statistical way. \citet{slater13} used the
Via Lactea simulations to show that a significant fraction of the dwarf galaxies
located between 300 and 1000 kpc from their host galaxy have made at least one
pericentric passage near a larger galaxy. However, the fraction of dwarfs that
have undergone two or more pericentric passages decreases sharply near 300 kpc.
This suggests that it is unlikely for And XXVIII to have undergone multiple
pericentric passages.

This presents a clear question for theories of dSph formation based on tidal
interactions: can a dwarf galaxy be completely transformed into a dSph with only a
single pericenter passage? Simulations of tidal stirring originally seemed to
indicate that the answer was no, and when dwarfs were placed on different orbits
it was only the ones with several ($\sim4-5$) pericenter passages that were
transformed into dSphs \citep{kazantzidis11a}. However, more recent simulations
that used cored dark matter profiles for the dwarfs suggest that multiple
pericenter passages might not be required. \citet{kazantzidis13} show that
dwarfs with very flat central dark matter profiles (inner power-law slopes of
0.2) can be transformed into pressure supported systems after only one or two
pericenter passages. This result is encouraging, but it also comes with the
consequence that cored dark matter profiles also tend to make the dwarfs
susceptible to complete destruction by tidal forces. In the simulations of
\citet{kazantzidis13}, five out of the seven dwarfs that were successfully
transformed into dSphs after only one or two pericenter passages were
subsequently destroyed. Taken together, these results indicate that rapid
formation of a dSph is indeed plausible, but there may only be a narrow range of
structural and orbital parameters compatible with such a process. Recent
proper motion measurements of the dSph Leo I support this picture even further,
as it appears to have had only one pericentric passage \citep{sohn13} yet is
unambiguously a dSph.

The properties of And XXVIII add an additional constraint that any tidal
transformation must not have been so strong as to completely mix the older and
younger stellar populations. A simple test case of this problem has been
explored by \citet{lokas12}, in which particles were divided into two
populations by their initial position inside or outside of the half light
radius. The dwarfs were then placed on reasonable orbits around a host galaxy,
and evolved for 10 Gyr. The resulting radial profiles of the two populations are
distinct in nearly all cases, with some variation depending on the initial
conditions of the orbit. These tests may be overly optimistic, since initial
differentiation into two populations is performed by such a sharp radius cut,
but the simulations illustrate the plausibility of a dwarf retaining spatially
distinct populations after tidal stirring.

An additional piece of the puzzle is provided by the metallicities. And XXVIII
and XXIX are both consistent with the luminosity-metallicity relation shown by
other Local Group satellites (see Section~\ref{spectra}). This implies that they
could not have been subject to substantial tidal \emph{stripping}, as this would
drive them off this relation by lowering the luminosity without substantially
altering their metallicities. This point is further reinforced by the similarity
of the luminosity-metallicity relation of both dSph and dIrr galaxies in the
Local Group \citep{kirby13}, making it unlikely that the measured
luminosity-metallicity relation itself is significantly altered by tidal
stripping. Whether or not more gentle tidal effects can
induce morphological transformation without altering the luminosity-metallicity
relation remains to be seen.

Taken together, the properties of And XXVIII and XXIX present a range of
challenges for detailed models of dwarf galaxy evolution to explain.
Particularly for And XXVIII, the wide separation and low mass of the system add
significant challenges to reproducing the gas-free spheroidal morphology with a
stellar population gradient, while there may be similar challenges for
explaining the apparent absence (or at least low-detectability) of such
gradients in And XXIX. Though plausible explanations have been shown to exist
for many of these features individually and under ideal conditions, whether the
combination of these conditions can be accurately reproduced in a simulation is
unknown. Further modeling of these types of systems is required before we can
understand the physical drivers of these observed features.

\acknowledgments

We thank the anonymous referee for their helpful comments which improved the
paper. This work was partially supported by NSF grant AST 1008342. Support for
EJT was provided by NASA through Hubble Fellowship grant \#51316.01 awarded by
the Space Telescope Science Institute, which is operated by the Association of
Universities for Research in Astronomy, Inc., for NASA, under contract NAS
5-26555.

Based on observations obtained at the Gemini Observatory, which is operated by
the Association of Universities for Research in Astronomy, Inc., under a
cooperative agreement with the NSF on behalf of the Gemini partnership: the
National Science Foundation (United States), the National Research Council
(Canada), CONICYT (Chile), the Australian Research Council (Australia),
Minist\'{e}rio da Ci\^{e}ncia, Tecnologia e Inova\c{c}\~{a}o (Brazil) and
Ministerio de Ciencia, Tecnolog\'{i}a e Innovaci\'{o}n Productiva (Argentina).

Some of the data presented herein were obtained at the W.M. Keck Observatory,
which is operated as a scientific partnership among the California Institute of
Technology, the University of California and the National Aeronautics and Space
Administration. The Observatory was made possible by the generous financial
support of the W.M. Keck Foundation.

The authors wish to recognize and acknowledge the very significant cultural role
and reverence that the summit of Mauna Kea has always had within the indigenous
Hawaiian community. We are most fortunate to have the opportunity to conduct
observations from this mountain.

{\it Facility:} \facility{Gemini:North (GMOS)} \facility{Keck:II (DEIMOS)}

\end{document}